# Flexible Asymmetrically Transparent Conductive Electrode based on Photonic Nanojet Arrays

D. Kislov[1], P. Voroshilov[2], A. Kadochkin[3,4], A. Veniaminov[2], V. Zakharov[2], V. V. Svetukhin[3], V. Bobrovs[5], O. Koval[6], I. Komendo[6], A. M. Azamov[7], A. Bolshakov[1,8,12,13], L. Dvoretckaia[8], A. Mozharov[8], A. Goltaev[2,8], V.Volkov[1], A.Arsenin[1], P. Ginzburg[9,10], I. Mukhin[8], A. Shalin[1,5,11]

[1] Moscow Center for Advanced Studies, Kulakova str. 20, Moscow, 123592, Russia
[2] ITMO University, Kronverkskii 49, St. Petersburg, 197198, Russia
[3] Scientific-Manufacturing Complex "Technological Centre", 124498 Zelenograd, Moscow, Russia
[4] Technological Research Institute, Ulyanovsk State University, 432017 Ulyanovsk, Russia
[5] Riga Technical University, Institute of Photonics, Electronics and Telecommunications, 1048 Riga, Latvia
[6] NRC "Kurchatov Institute", 123182, Moscow, Russia
[7] Nukus Innovation Institute, Republic of Uzbekistan, Republic of Karakalpakstan, Murtazabiy 28, Khojaly district
[8] Alferov University, Khlopina 8/3, Saint Petersburg 194021, Russia
[9] Department of Electrical Engineering, Tel Aviv University, Ramat Aviv, Tel Aviv 69978, Israel
[10] Light-Matter Interaction Centre, Tel Aviv University, Tel Aviv, 69978, Israel
[11] Faculty of Physics, M. V. Lomonosov Moscow State University, 119991 Moscow, Russia
[12] Laboratory of Advanced Functional Materials, Yerevan State University, Yerevan, 0025, Armenia
[13] Faculty of Physics, St. Petersburg State University, Universitetskaya Emb. 13B, St. Petersburg 199034, Russia

**Abstract**

Flexible transparent electrodes, encompassing the combination of optical transparency and electrical conductivity, empower numerous optoelectronic applications. While the main efforts nowadays concentrate on developing wire meshes and conductive oxides, those technologies are still in a quest to find a balance between price, performance, and versatility. Here we propose a new platform, encompassing the advantages of nanophotonic design and roll-to-roll large-scale lithography fabrication tools, granting an ultimate balance between optical, electrical, and mechanical properties. The design is based on an array of silica microspheres deposited on a patterned thin aluminum film attached to a flexible polymer matrix. Microspheres are designed to squeeze 80% light through nanoscale apertures with the aid of the photonic nanojet effect given the light impinges the structure from the top. The photonic structure blocks the transmission for the backpropagation direction thus granting the device with the high 5-fold level of asymmetry. The patterned layer demonstrates a remarkable 2.8 Ω/sq sheet resistance comparable to that of a continuous metal layer. The high conductivity is shown to be maintained after a repeatable application of strain on the flexible electrode. The technical specifications of the demonstrated transparent electrode establish it as a viable option for integrating into advanced optoelectronic devices such as solar cells, touchscreens, and organic light-emitting diodes to name a few. Its notable capacity to optimize light transmittance while ensuring consistent electrical performance, alongside its mechanical flexibility, makes the demonstrated device an essential component for applications, where such attributes are critically required.

# 1. Introduction

Flexible transparent conductive coatings (TCC) have become essential components, widely used across modern technologies, where both optical and electronic functions must be superimposed on the same area of a device. Those include and are not limited to touch screens, light-emitting diodes, photodetectors, solar cells, smart windows in the automotive industry, and many others. High-grade TCCs accommodate several key parameters simultaneously, including (i) low sheet resistance, (ii) high optical transparency, (iii) stability (including chemical), (iv) mechanical flexibility, and (v) conformity for a device form factor [1-4].

Nowadays, indium-tin-oxide (ITO) is the prevailing material platform, used for TCCs. This conductive oxide has excellent transparency, along with a relatively low ~ 10 $\Omega$/sq sheet resistance [5]. However, ITO layers are rather fragile and cannot be straightforwardly applied to flexible devices. Furthermore, ITO fabrication requires employing high-temperature deposition or annealing processes, which further challenges its integration within complex large-scale devices [5],[6]. Considering the natural rarity of indium [7],[8] alongside the previously mentioned technological challenges, it calls for developing new strategies for TCC architectures [6]. Alternative conductive oxides, which can be used for TCCs, include Al-doped ZnO (AZO), Ga-doped ZnO (GZO), and F-doped $SnO_2$ (FTO) to name the key ones [9]. Carbon-based materials, including single-walled and multi-walled carbon nanotubes [10-16], graphene [10, 11, 17, 18], reduced graphene oxide (RGO) [11,19, 20], and graphene quantum dots (GQD) [21], being intensively explored nowadays, are promising candidates as well. Alternative platforms to mention are conductive polymers including PEDOT:PSS [22-24], metal nanowire networks [25, 26] (including coatings [27, 28]), metal-dielectric layered composites [29-36], widely used traditional patterned metal layers [37, 38, 39], and related hybrid nanocomposites [20, 40-43]. One of the main remaining challenges in the field is to balance between mechanical flexibility of electrodes optimizing the tradeoff between optical transparency and sheet resistance [38].

Here we explore an architecture, based on micro-lens arrays, superimposed with a patterned metal film on flexible substrate. This layout allows to benefit from both high optical transparency and high conductivity of a large-scale layer, potentially reaching 10s of $cm^2$ areas. We demonstrate that the sphere array focuses and guides light through holes in a metal film utilizing the photonic nanojet (PNj) phenomenon [44]. The lensing effect of light transmission through a small aperture is demonstrated to grant an asymmetric optical transmission, which is among the demanded functions in application to smart windows, solar cells, and several others. Our test samples encompass 3 inches (7-8cm) across and are implemented on flexible substrates, paving the way for the implementation of conformal transparent electrodes. In terms of mechanical and electrical properties, our device will be demonstrated to reach performances of continuous thin metal films [45] while surpassing them in optical transparency over the entire visible range.

The manuscript is organized as follows: first, we reviewed the design of the TCC, concentrating on the main parameters of the system and the operation principle (section 2); section 3 discusses the main results of the work: the results of optimization of the metal patterned layer and numerical simulation of the transmission of TCC are presented in section 3.1; Next, the method of fabricating TCC and the results of measurements of the transmission and resistance of a patterned metal film are presented (sections 3.2 - 3.3); in section 3.4, TCC's optical asymmetry is considered within the Lorentz reciprocity principle; further in section 3.5, the optical asymmetry of the TCC is demonstrated based on transmission measurements; at the same section, mechanical flexibility of the TCC is demonstrated. Finally, the potential applications of the presented TCC are discussed before the Conclusion.

## 2. TCC Architecture

The proposed TCC architecture appears in Figure 1a. The first stage of optimization aims to enhance the optical transmission of the film. The general engineering tradeoff is made between the transparent area and electrically conducting optically opaque metal network. While extraordinary optical transmission empowered by plasmonic phenomena in hole arrays was demonstrated, it still grants ~26% transparency (at 2.01 Ω/sq sheet resistance) [39], which is less practical for TCCs. To enhance the transmission through the hole array, we propose to decorate it with dielectric (e.g. glass) microspheres - the layout of the microsphere array appears in Figure 1b. Dielectric microspheres are known to form PNj – light focusing in close vicinity to the particle surface [44]. Figure 1c demonstrates the effect of the PNj narrow beam, passing through opaque metal aperture (technical details of the model will follow). It can be seen that the effective area of the hole is increased several fold owing to the light collection and collimation effect provided by microsphere.

In general, the field distribution in a PNj depends on refractive indexes and the overall geometry and thus can be tailored accordingly. With an outlook on the subsequent experimental realization based on colloidal self-organization, hexagonal close-packing (*hcp*) in monolayer and 90% occupation of the surface will be assumed. The fractional transmission thus can be estimated with the aid of geometrical optics (normal incidence):

$$\frac{S_1}{S_2} = \frac{2\pi R^2}{4\sqrt{3} R^2} = \frac{\pi}{2\sqrt{3}} = 0.907, \tag{1}$$

where $S_1$ – area occupied by hexagonal close-packed spheres, $S_2$ – unit cell area, and $R$ – sphere radius. Given that PNj are properly designed, ~90% transmission can be achieved in the case of lossless structures. Note that *hcp* has advantages as it allows approaching an almost perfect aperture efficiency, compared to other packing. Notably, the spheres exhibit self-organization on the surface into precisely such a lattice. Considering this design, we will demonstrate: (a) flexibility, (b) low sheet resistance due to a relatively large area of metal coverage, and (c) transmittance close to 90% (taking into account the reflection from the microsphere layer) in a wide spectral range, due to the non-resonant PNj that provides light propagation without reflection and absorption. Additionally, when the device is illuminated from the bottom side (from the flexible substrate), a much stronger reflection is expected leading to a decrease in the backward transmittance. This asymmetry effect is important from both fundamental and applied standpoints and will be discussed below.

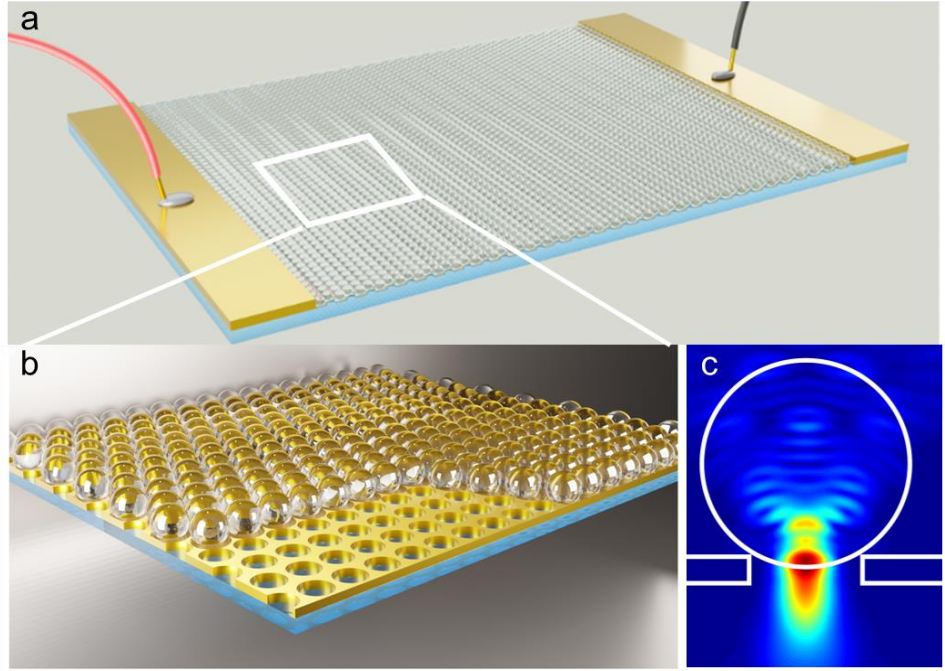

**Figure 1**. Flexible transparent conductive metamaterial based on a micro-lens array and patterned metal layer. (a) Schematics of the macroscopic array with electrodes. (b) Zoomed fragment of a surface, a few microspheres were removed to show the patterned metal layer design. (c) The operational principle of PNj, granting efficient light transmission through a metallic aperture – numerical analysis.

## 3. Results and discussion

*3.1 Optimization of the metal film*

The finite element method (FEM), implemented in the COMSOL Multiphysics software package [46] was used for numerical analyses. Infinite periodic structure with the unit cell presented in Figure 2a is considered. Aluminum (Al) and silica (for the film and spheres, respectively) were chosen for the material platform. Al was chosen due to its high electrical conductivity, cost-effectiveness, and fabrication feasibility.

The span of microspheres radii ($R$) in the range of 550 – 850 nm was considered [47]. The use of larger spheres will lead to a partial loss of the signal due to reflection from the metal. The size of the holes ($r$) is used as an additional optimization parameter. In the first set of optimizations, the metal film thickness ($h$) is taken to be 180 nm. This parameter will be discussed in the next section. The structure lies on a flexible poly methyl methacrylate (PMMA) substrate and is surrounded by air.

First, we calculate the transmittance as a function of the size of silica spheres and holes in the metal film. For getting a 2D colormap, the transmittance was averaged over the 400 - 800 nm range (see Figure 2b). While the transmittance weakly depends on the microsphere radius, it has a strong response to the hole size. The transmittance drops to 40% for radii below 200 nm, while it is expected to overcome 80% with larger than 400 nm holes.

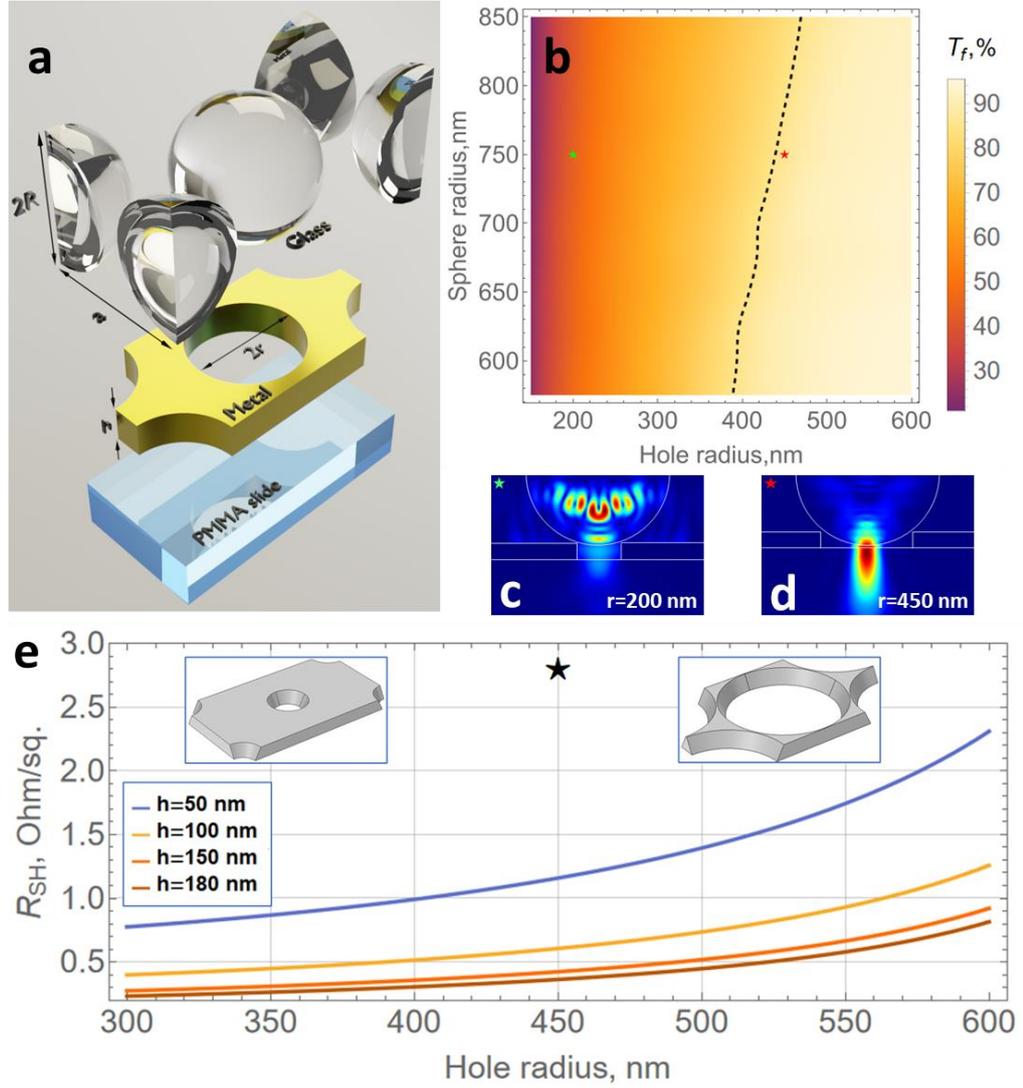

**Figure 2.** The device transmittance and resistivity – numerical analysis. (a) Schematic of the unit cell: patterned Al film (thickness $h$) with cylindrical holes (radius $r$) in a hexagonal lattice with pitch $a$. Silica microspheres of radius $R$ over the holes, $2R = a$. The structure is placed on a flexible PMMA substrate. (b) Colormap of the transmittance (averaged over 400 - 800 nm) as a function of the hole and sphere radii. Here the dotted curve shows a transmittance value of 85%. (c, d) Electric field intensity distribution across a unit cell. The plane wave propagates from top to bottom (forward transmission). Parameters: $a = 2R = 1500$ nm, Al film thickness $h = 180$ nm, hole radius $r = 400$ nm (labeled as green ★ on the map (b)) and (d) same parameters, but $r = 450$ nm (labeled as red ★ on the map b)). Wavelength – 500nm. (e) The calculated sheet resistance of the patterned Al film (pitch $a = 1500$ nm) as a function of the hole radius for different values of the film thickness $h$, experimentally obtained sheet resistance labeled as (★).

To visualize the light propagation mechanism, the field distributions for 200 nm and 450 nm holes were calculated. As an example of a n unsuccessful design, we plot the field distribution for the same sphere and 200 nm hole. The film strongly scatters the light and the transmission drops (see Fig. 2c). On the other hand, when the hole radius is increased to 450 nm, as expected, the microsphere collimates a plane wave into a narrow PNj (see Figure 2d).

In summary, $R$=750 nm spheres and $r = 450$ nm holes promote 85% transmittance. A small tolerance on the spheres' size provides fabrication flexibility [47]. In the following analysis, those parameters will be chosen.

*3.2. Optimization of Al film thickness*

The sheet resistance of the patterned Al film as a function of the hole size and film thickness is calculated next. Figure 2d demonstrates that the PNj promoted by an *R=750* nm sphere supports efficient light propagation through the *r=450* nm hole in a 180 nm thick Al film. The film thickness thus should not exceed 180 nm, otherwise an excessive light scattering is observed. The sheet resistance for thickness over 150 nm does not exceed 0.5 Ω/sq for the 450 nm holes and 1500 nm period. This is a rather low value assuming 0.28 Ω/sq resistance of a 100 nm thick continuous Al film. When the thickness decreases to 50 nm, the sheet resistance goes above 1.2 Ω/sq.

180 nm thick films will be explored hereinafter. From the experimental perspective, this thickness allows applying high-quality coating, suppresses defects during the metal surfacing, and also ensures structure uniformity. In addition, an increase in the metal film thickness may adversely affect the mechanical flexibility of the device.

*3.3 Fabrication and Characterization*

TCC was fabricated using a feasible, scalable, and cheap combination of vacuum thermal deposition, optical lithography through an ordered array of spin-deposited microspheres, and wet metal etching. Figure 3a depicts a schematic of a fabrication protocol. 180 nm thick Al film is deposited on a transparent flexible substrate (PMMA) using vacuum thermal evaporation (step 2 in Figure 3a). Next, AZ1505 positive optical resist (100 nm thickness) and a monolayer of 750 nm radius fused silica microspheres (Microparticles GmbH) were successively spin-coated over Al film (step 3). These self-organized microspheres were arranged in a close-packed *hcp* and used as PNj sources for the patterning of the resist according to the protocol reported previously [47]. The resist was exposed through these microspheres using the i-line of a mercury lamp (365 nm) (step 4). During the subsequent resist development, this first microsphere layer was washed off and the patterned resist layer on the metal surface with submicron holes with *hcp* lattice was formed. This resist layer served as a mask during the consequent wet etching of the metal film (5). After the etching and resist removal, a patterned metal layer with *hcp* submicron holes was obtained. At the final technological stage, a second monolayer of silica microspheres of the same diameter was deposited via spin coating. These microspheres predominantly occupied holes in the metal film (step 6). It is worth emphasizing that all the technological steps outlined here can be seamlessly scaled up to accommodate large-diameter samples. More specifically, the deposition of the microsphere monolayer, which constitutes the most crucial stage, can be flawlessly executed on substrates measuring 3 inches or even larger.

The optical image of the fabricated sample is demonstrated in Figure 3b together with the scanning electron microscopy (SEM) image of the *hcp-ordered* spheres (on inset). Figure 3c and d depicts the enlarged SEM images of the structure with highlighted (red dashes) unoccupied holes in the metal film. The SEM image in Figure 3d reveals an individual unoccupied hole, exhibiting sidewalls that are subtly false-colored. These sidewalls measure a minimum radius of 450 nm, which aligns with the desired specifications.

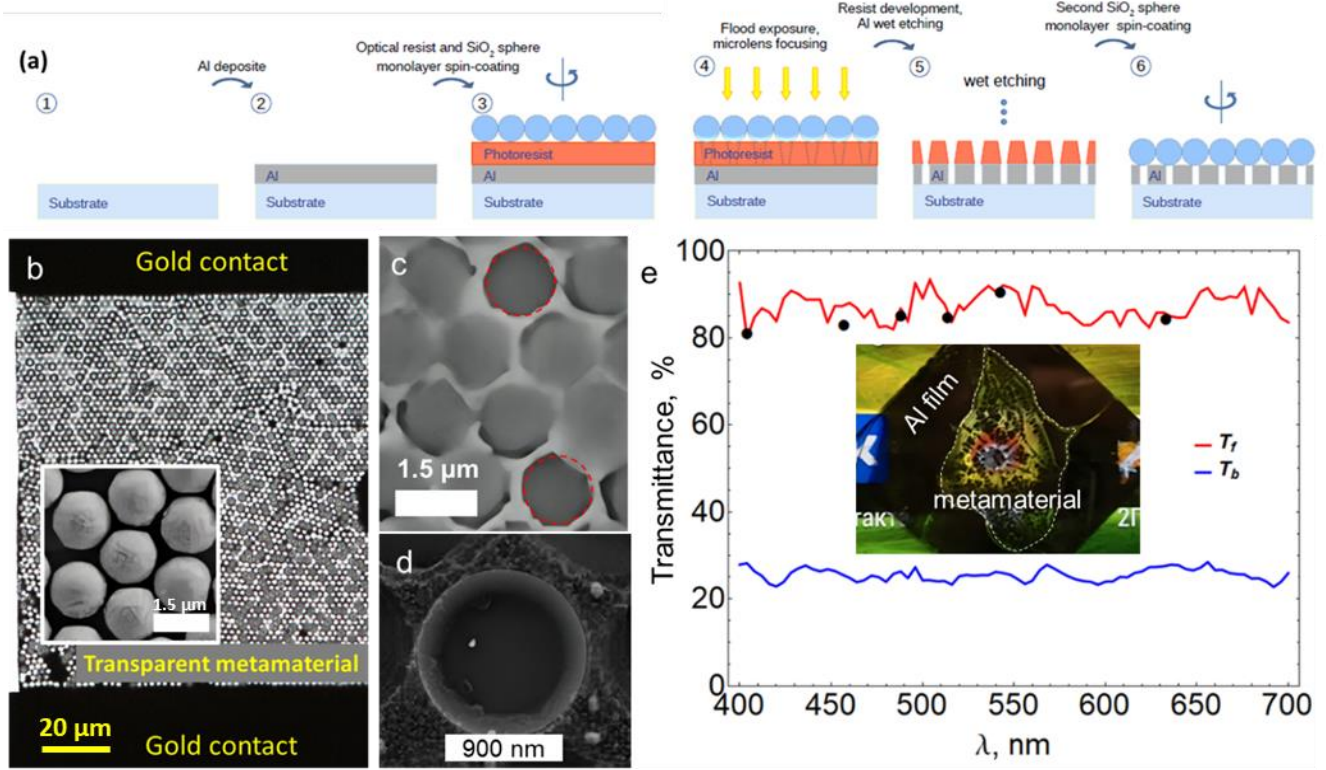

**Figure 3.** TCC fabrication and characterization. (a) Main fabrication steps. (b) Optical image of the fabricated sample (180 nm thick patterned Al film, holes $r$ = 450 nm with silica spheres $R$ = 750 nm) with Au contact pads, inset–zoomed SEM image of the surface. (c) Zoomed SEM image demonstrating unoccupied holes - highlighted with red dashed circles. (d) SEM image of an individual hole in the patterned Al film. (e) Forward (red) and backward (blue) transmittance spectra of the layer, dots - experimentally measured forward transmittance, inset - optical image of the sample, and unmodified continuous Al film placed over the operating smartphone screen showing the high transparency of the TCC.

To characterize the optical properties of the fabricated structure, we first evaluate numerically its' transmittance spectrum. The results of the modeling presented in Figure 3e show the calculated forward (red curve) and backward (blue curve) transmittance spectra demonstrating high theoretical forward transparency exceeding 80% and backward transmission below 30% throughout the visible range.

An experimental study of sample transparency was carried out using an LSM710 scanning confocal system based on an Axio Imager Z1 upright microscope (Carl Zeiss). Using the microscope, we measured the local transmittance of various microscopic areas of the sample to study inhomogeneities and design defects that arose during the fabrication. During the measurements, the sample was illuminated consecutively with 405, 458, 488, 514, 543, and 633 nm CW lasers. Our study examined the transmission of a weakly diverging light beam, produced by an optical system with a numerical aperture (NA) of 0.2, as it passed in the forward direction from the side of a spherical microparticles array and in the reverse direction from the metal film side. The intensity was measured using a photomultiplier. The transmission value was then evaluated as the ratio of the intensity of the light passed through the sample and a bare PMMA substrate. The obtained experimental data shown in Figure 3e with dots demonstrate more than 80% high forward transmittance in agreement with the numerical data. Weak spectral dependence of the transmittance over the entire visible range reflects the non-resonant nature of the device. Note that in the case of a metal structure, plasmon resonances can occur slightly increasing transmission through the holes [48]. However, this effect is rather narrowband. In this case, the transmittance fell significantly short of 80%, while in our case, the forward transmittance is 90% under 500 nm illumination. Inset in Figure 3e shows an optical image of the sample on a smartphone screen demonstrating the transparency. In contrast, part of the sample does not exhibit transparency because during the fabrication Al film was patterned partially. Additionally, the spatial distribution of the forward transmittance over the sample surface was studied. It was found that 3 levels of transmittance correspond to defect-free sections; sections of inaccurate stacking, where the microspheres are displaced relative to the holes, and sections of the patterned metal layer not covered with microspheres (see details in Supplementary Information S1).

To measure the film resistance, gold contact pads were fabricated - Figure 3b. Electrical measurements were carried out according to the four-point method using a probe station. The specific surface resistance was found at 2.8 Ω/sq. Note that the patterning of a continuous metal layer reduced its initial conductivity by no more than 2 times. This experimental sheet resistance value labeled as (★) in Figure 2e is an order of magnitude higher than the theoretically calculated, which can be explained for several reasons. First, the dispersion in the hole diameter over the entire film, due to patterning inhomogeneity, is related to the wet etching method [49]. Second, wet etching can violate the film integrity in the near-boundary etching regions, which should reduce the overall layer conductivity. Third, the sputtered gold contact pads can also induce additional resistance [50]. In addition, in the theoretical calculations, we used resistivity value corresponding to pure bulk Al, without taking into account impurities and oxidation.

*3.4. Optical transmission asymmetry*

The earlier sections showed that the sample's transmittance stays uniform throughout the entire visible spectrum. In this part, we explore the scattering effects caused by the geometry, focusing on how the signal collection efficiency varies with the numerical aperture of the objective used. The aperture of the objective determines the solid angle of the incoming light.

Linear time-invariant nonmagnetic structures obey the Lorentz reciprocity principle [51]. The device reported here falls under this category. However, the demonstration of asymmetric optical transmission does not violate the reciprocity principle given the energy passing through the device is collected with a considerably large (NA). Silica spheres focus light into the small aperture, from which it diffracts on the opposite side of the metal film. Light, impinging the aperture from the opposite direction does not experience focusing, thus the transmission is estimated with the fractional area of apertures in the metal film. Moderate plasmonic corrections affect this estimate and are considered with the full-wave numerical simulation.

We analyze the transmission asymmetry by comparing the ratio of forward to backward transmittance. Considering the geometric parameters of the system, we plot the transmittance and asymmetry maps (refer to Figures 4a and c, correspondingly). When we increase the microsphere radius, the forward transmittance remains virtually unchanged, while the backward transmittance gradually decreases (see Figure 2b and Figure 4a, correspondingly). This decrease in backward transmittance is attributed to the reduction in the relative hole area as the structure lattice parameter (determined by the spheres' diameter) increases, assuming a fixed hole radius. Conversely, the reduction in the relative area of the holes is not substantial for forward transmittance, as the spheres capture the incident light. The proportion of their geometric cross-section to the total area, as described in Eq. 1, remains unchanged. The backward transmittance is slightly lower, on average, than the relative hole area (e.g., for hole $r =$ 400 nm and sphere $R = 750$ nm, the backward transmittance is approximately 20% while a relative hole area is slightly below 26%). This effect arises from the sub-wavelength hole sizes. Furthermore, when light propagates from the bottom (metal film side), the field experiences a significant reflection, rendering the energy transmittance through the hole much weaker (see Figure 4b). Therefore, we can infer that as the sphere radius increases, the transmission asymmetry in this system also goes up (see Figure 4c).

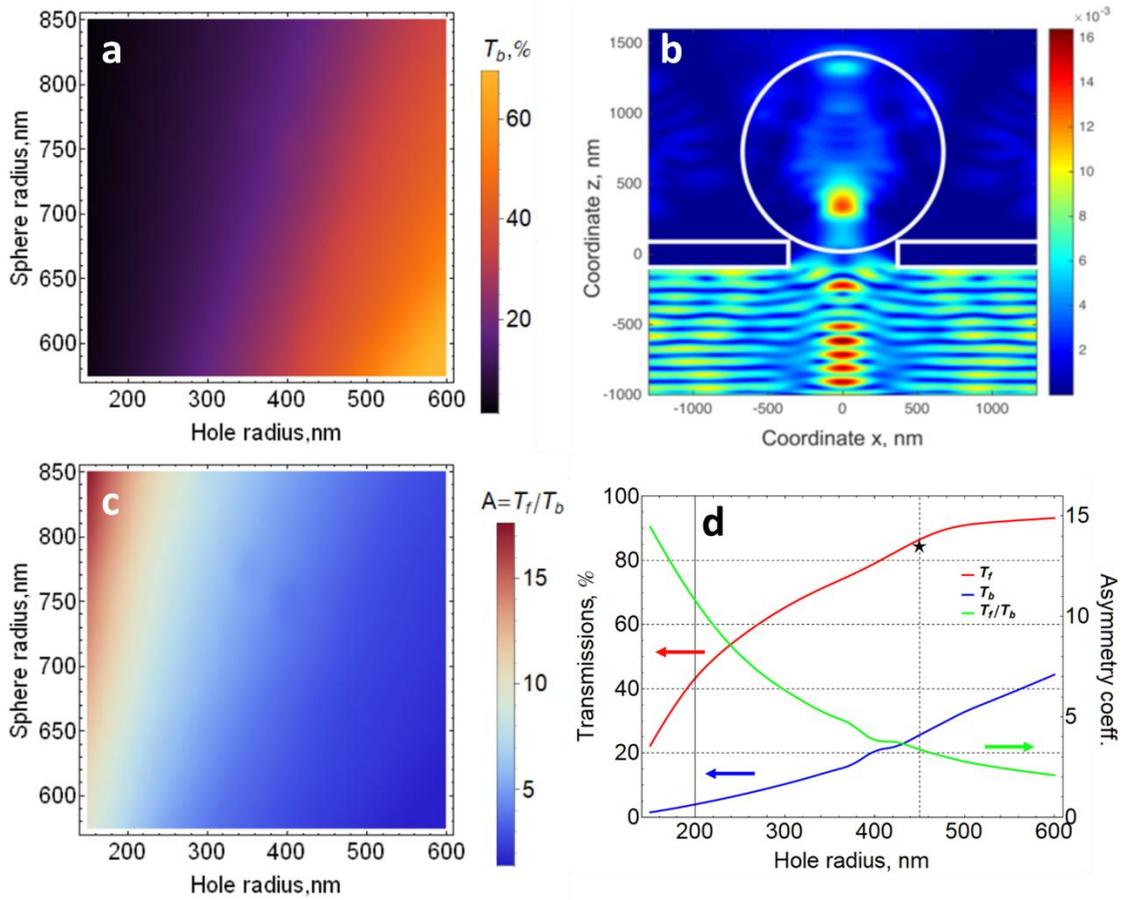

**Figure 4.** TCC optical asymmetry (Al film thickness $h = 180$ nm). (a) Backward transmittance, averaged over the optical 400 - 800 nm wavelength range. (b) Electric field distribution for a plane wave incident from the direction of the metal film (backward direction). Parameters: $a = 2R = 1500$ nm, hole radius $r = 450$ nm, PMMA substrate, excitation wavelength - 500 nm. (c) TCC asymmetry factor (the ratio of forward to backward transmittance), as the function of the system geometry. (d) The forward transmittance ($T_f$, red curve), backward transmittance ($T_b$, blue curve), and asymmetry factor ($T_f / T_b$, green curve) as a function of a hole radius. Microspheres radius - 750 nm and a film thickness -180 nm. Wavelength-averaged experimentally obtained forward transmittance is labeled as (★).

Figure 4d illustrates the simulation results on the forward and backward transmittance as well as the asymmetry averaged across the entire optical range. These results correspond to the specific experimental parameters adopted in this study (Al film thickness – 180 nm, $R=750$ nm microspheres). The backward transmittance exhibits a relatively slow increase as the hole radius increases, ranging from 1.5% at $r = 150$ nm to 44.5% at $r = 600$ nm. Figure 4d also shows that the forward transmittance curve has a structure. Within the range of hole radii from $r = 150$ nm to $r = 250$ nm, there is a rapid increase in transmittance from 22% to 56%. Subsequently, as the hole radius ranges from $r = 250$ nm to $r = 500$ nm, the increase in transmittance slows down. Upon reaching a hole radius of 500 nm, the curve levels off, resulting in an average transmittance value of 91 – 93%. This value is approximately equivalent to the average microsphere cross-sectional area and corresponds to the transparency of the substrate. The asymmetry coefficient is maximized at $r = 150$ nm, with a forward transmittance of 22% and an associated value of 14.5. However, the asymmetry coefficient decreases thereafter. At $r = 600$ nm, the asymmetry coefficient equals 2, while the backward transmittance increases to 44.5%. Thus, by varying the hole diameter, it is possible to implement the mode of maximum transmittance with a simultaneous moderate resistance and significant optical asymmetry (~2). Thus, by modifying the parameters, it's feasible to significantly change the optical and electrical properties of the structure over a wide range. The variability of forward to backward transmittance ratio empowers a wide range of practical applications: from screens and solar cell coating to tinted conductive films. Note that the sheet resistance in the specified range of the optimized parameters is quite small and does not exceed 1 Ω/sq, which is especially low value at a transmittance of 90%. In comparison, the most widely used commercial ITO has $R_S$ = 10 Ω/sq with transmittance not exceeding 85% [5] (see Table 1).

*3.5 Transmittance and Bending Sustainability*

As the aperture of the optical system is enlarged, facilitating the collection of light transmitted through the sample, the measured light intensity, as well as the transmittance, increases in both forward and backward directions (Fig. 5a). Notably, the backward transmittance saturates below 40% while for the forward this value exceeds 80%. When the numerical aperture is small, the transmittances in both directions are similar. As noted above, zero-order diffraction is consistent with the Lorentz reciprocity principle and gives the same light transmittance for both forward and backward directions [51]. An additional transmission is provided by the diffraction side lobes having a certain opening angle. Indeed, the impact of beams deflected at a large angle from the initial direction of the photodetector, due to the use of lenses with a large aperture, makes it possible to observe a significant optical asymmetry (~2.2).

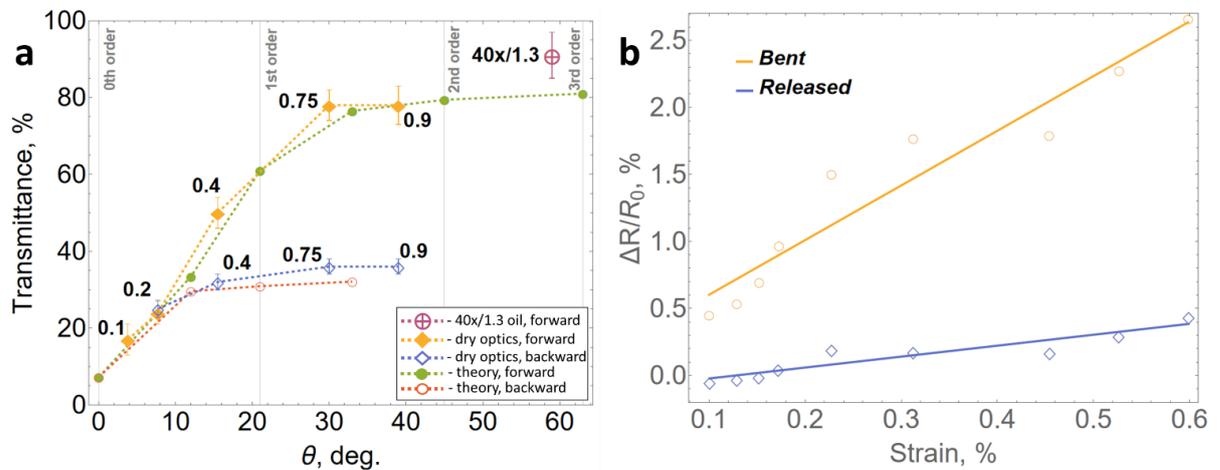

**Figure 5**. Transmittance and bending sustainability of the considered metamaterial. (a) Angular dependence of light intensity passing through an ordered section of the metamaterial: theoretical (corresponding to the calculated intensities of 0 - 3rd diffraction orders in the forward direction) and measured under 405 nm illumination for forward and backward transmittance directions. The point indicated by a circle with a tick was obtained by measuring the transmittance in the forward direction using an immersion objective. The numbers on the graph correspond to the numerical aperture of the objectives. (b) The change in relative electrical resistance upon application of mechanical strain. $R_0$ – initial sample's resistance.

The measurements were performed using a 405 nm laser source. The experimentally obtained forward transmittance is close to the calculated values of the diffraction efficiency in 0–3 orders, while the latter can be taken from the electrode substrate only when an immersion liquid is used between the substrate and objective. In this case, the total efficiency of experimentally obtained four diffraction orders was 92%. However, due to significant errors that arise when measuring with an immersion objective, this transmittance value should be treated with caution as the maximum possible value calculated in the geometric optic framework. At the same time, as seen in Figure 5a the forward and backward transmittance angular dependences tend to value of 80% and 36%, respectively, with increasing numerical aperture. This agrees well with the experimentally measured and calculated data on the integrated transmittance under illumination with a 405 nm wavelength (see Figure 3c). The spatial distribution maps of the light intensity passing through the electrode in the forward and backward directions and collected by lenses with different apertures, as well as the diffraction pattern are shown in the Supplementary Information Figure S2, and S3.

Table 1. Comparative characteristics of various TCCs

| Material | Sheet resistance, Ω/sq | Transmittance | Flexibility | Advantages | Disadvantages | Ref. |
|---|---|---|---|---|---|---|
| ITO | >14 | 85-90% | - | Large area, Industrial use | Fragility, high cost, limited stock of raw Indium, high temperature required during sputtering or post-annealing | [5], [6] |
| ZnO:Al (AZO), ZnO:Ga (GZO), SnO2:F (FTO) | >100 | 85-90% | - | Possibility of large-area coating | High sheet resistance, rough surface impairing electrical contact with subsequent layers | [9] |
| Graphene | 219 to $1.35 \times 10^5$ | 70 - 96,5 % | + | Possibility of large-area coating | High sheet resistance, scaling technology difficulty over a large area | [10], [11],[17], 18] |
| RGO | $10^3$-$10^6$ | 57 - 87% | + | Possibility of large-area coating | High sheet resistance, scaling technology difficulty over a large area | [10, 11 ,19, 20] |
| Graphene quantum dots | 290 | 88 - 90% | + | Possibility of large-area coating | High sheet resistance, scaling technology difficulty over a large area | [21] |
| CNTs | 10-1000 – Single-walled; 180-250 - Multi-walled | 70 - 95%; 38 - 81% | + | Possibility of large area coating, industry use | Low optical transparency, scaling technology difficulty over a large area | [10-16] |
| PEDOT:PSS | >100 | 90% | + | Possibility of large area coating, industry use | low free paths of charge carriers, the necessity of the use of additional metal networks | [22-24] |
| Metal nanowire networks | 4-8 | 90% | + | Possibility of large area coating, industry use | High surface roughness, not suitable for charge carrier collecting into the materials with low mean free paths | [25, 26] |
| Metal NWs | 10-100 | 80 - 90% | + | Possibility of large area coating, industry use | Relatively high surface roughness, coating inhomogeneity | [27, 28] |
| Patterned metal film | 2; 10; 0.93 | 26%; 80%; 90% | + | Possibility of large area coating, industrial use | Low optical transmittance | [37, 38, 39] |
| Related hybrid nanocomposites | > 13.9 | Up to 87,5% | + | Possibility of large area coating, industrial use | Fragility, different transmittance values over the electrode volume, coating inhomogeneity | [17, 20, 40-43] |
| layered dielectric-metal-dielectric structures | 5,75 | 91,6 % under 550 nm | - | Possibility of large area coating, industrial use | Technological manufacturing complexity/difficulty, narrow transparency spectral range | [29-36, 38] |
| Flexible Asymmetrically Transparent Conductive Electrode based on Photonic Nanojet Arrays | 2,8 | 80 - 90% | + | Possibility of large area coating, industrial use | - | This work |

A practical merit of TCC is its sustainability under deformation. Figure 5b shows the measured resistance change upon the bending action. The sample was transferred on a stretchable surface undergoing strain, and the layer conductivity was measured. At the end, the sample was then transferred back to a flat surface, and its conductivity was measured again after strain relaxation. A series of measurements using surfaces with different curvatures was carried out, the maximum resulting mechanical strain corresponded to 0.6%, and the minimum surface curvature radius was about 10 mm, the measurement technique is described in more detail in our previous work [16]. Figure 5b demonstrates that an increase in mechanical strain leads to the conductivity decrease. When the strain is released, the resistance is almost restored. After an experiment series (more than 100 cycles), the overall layer conductivity deteriorates by less than 0.5%, which confirms the applicability of the proposed metamaterial for flexible applications.

Finally, we present a summary of the most recent advancements in competitive and promising TCC materials (see Table 1). This study reveals the remarkable elasticity, pristine conductivity, and transparency of the reported TCC. Moreover, it identifies a distinctive optical asymmetry property, which further distinguishes this platform from other transparent conductive materials.

# 4  Conclusion

In summary, we presented the design, fabrication, and analysis of a flexible transparent conductive electrode, tailored for optoelectronic applications. The new architecture is composed of a matrix of silica microspheres uniformly distributed over a patterned thin metal film, which is itself situated on top of a flexible substrate (PMMA). The electrode exhibits remarkable electrical and optical properties, demonstrating a 2.8 $\Omega$/sq sheet resistance comparable to that of a continuous metal layer, while maintaining a high optical transmittance of over 80%. Furthermore, the electrode exhibits remarkable mechanical resilience, maintaining consistent resistance with minimal variance even under significant bending strains, and demonstrating negligible changes in resistance post-strain relief. Additionally, the system's performance remains largely unaffected by minor variations in the dimensions of the holes and spheres (up to 10%), underscoring its robustness. The experimentally validated architecture presented in this report shows promise for various applications, particularly in optoelectronics. It addresses the growing demand for integrating conductive components into devices while preserving both transparency and flexibility.

**Acknowledgments**
 Ministry of Science and Higher Education of the Russian Federation (Agreement No. 075-15-2022-1150); The calculations of the optical transmittance are partially supported by the Russian Science Foundation grant №23-72-00037. A.G. thanks the Russian Federation (Agreement No. 21-79-10202) for sample gabrication.  P.G. acknowledges Science Forefront (Israel), project 0006764. V.B. acknowledges Latvian Council of Science, project "DNSSN" (project No. lzp-2021/1-0048). The project started in 2019, there is no joint funding between the teams.

### ASSOCIATED CONTENT

**Supporting Information**.
This material is available in the next .docx file.